# A Process Model for Crowdsourcing: Insights from the Literature on Implementation

Alireza Amrollahi
School of Information and Communication Technology
Griffith University
Gold Coast, Queensland, Australia
Email: alireza.amrollahi@griffithuni.edu.au

## Abstract

The purpose of the current study is to systematically review the crowdsourcing literature, extract the activities which have been cited, and synthesise these activities into a general process model. For this to happen, we reviewed the related literature on crowdsourcing methods as well as relevant case studies and extracted the activities which they referred to as part of crowdsourcing projects. The systematic review of the related literature and an in-depth analysis of the steps in those papers were followed by a synthesis of the extracted activities resulting in an eleven-phase process model. This process model covers all of the activities suggested by the literature. This paper then briefly discusses activities in each phase and concludes with a number of implications for both academics and practitioners.

## 1 Introduction

Although the term crowdsourcing has recently entered the Information Systems (IS) literature, the use of collective intelligence of large number of people for solving business and academic problems has been largely subject of attention throughout history (Leimeister, 2010; Pedersen et al., 2013). However, after introduction of the term by Howe (2006b) as a new sourcing approach, it attracted increased attention from both academy and practice perspectives.

Crowdsourcing approach has been used for solving several diverse problems up to now. Seltzer and Mahmoudi (2013) have reviewed 24 crowdsourcing platforms for a variety of application such as: business, city planning, policy development, and event outreach. Crowdsourcing.org has also listed 2670 sites in 45 languages which shows an increase of more than %100 between 2011 and 2013 (Tarrell et al., 2013). The Amazon Mechanical Turk which is one of the most famous sites in the world, has more than 571,000 tasks on April 2014. IStock is another successful platform which is dedicated to the photography industry. This platform is purchased by Getty Images for $50 million in 2006 (Howe, 2006b) and its revenue in 2008 was approximately $163 million (Pickerell, 2012). Many businesses also have used the model to improve their products and services. "Idea storm" of Dell for example is used for submission of ideas about new products (Poetz and Schreier, 2012) and already contains more than 20,000 ideas. The approach has recently used for even high level decision making and strategic planning (Amrollahi et al., 2014; Stieger et al., 2012).

Moreover, as stated by Paolacci et al. (2010), most of the users in crowdsourcing platforms, contribute for reasons other than monetary motivations and this model is widely used for academic, scientific and non-commercial purposes: Foldit is a famous example of the use of the model in scientific problems. It has been developed by David Baker's lab at the University of Washington, to apply the crowdsourcing model to the protein structure prediction. This approach helped the lab to resolve a problem which was unsolved for scientist for decades (Cooper et al., 2010; Graber and Graber, 2013). Ranard et al. (2013) has also mentioned 21 cases of using crowdsourcing model in health, medicine, psychology and human behaviour areas.

Along with advances in the application of crowdsourcing model in practice, researchers have also paid a great attention to this area. An analysis of the 15 top IS journals and conferences by Tarrell et al. (2013) resulted in 135 articles which studied different aspects of this model after 2006. Processes for utilisation of the crowdsourcing is one of these research areas which has been defined by Pedersen et al. (2013, p. 581) as "the design of a step-by-step plan of action for solving a crowdsourcing problem."

Although various studies have mentioned different activities for implementing the crowdsourcing approach, most of the work is context based or ad hoc and no comprehensive approach has been developed to this date. For this reason, the crowdsourcing literature lacks a comprehensive guideline through which practitioners can initiate and manage their crowdsourcing projects. This shortcoming motivated the authors to perform a comprehensive review of the literature and synthesise the activities which have been proposed in previous studies into a generic process model. The relevant research question for the current study is:

RQ. Which activities or phases have been introduced as part of the crowdsourcing process model?

The results of this study may help future research by pointing out the gaps in the current body of literature. The developed process model may also help practitioners to compare different phases and activities, and select or customise them based on the context in which they want to use the crowdsourcing model.

## 2 Research Background

Howe (2006b) first defined the term crowdsourcing as: "the act of taking a job traditionally performed by a designated agent (usually an employee) and outsourcing it to an undefined generally large group of people in an open call (Howe, 2006a)." However in subsequent research and practice in the field, two dimensions of this definition were questioned: first, organisational stakeholders and employees have been involved in the process and second, organisations have started to select the crowd for participation.

To address these modifications, upcoming research has provided different definitions. Brabham (2009, p. 252) for example defined crowdsourcing as "a legitimate, complex problem-solving model, more than merely a new format for holding contests and awarding prizes . . . It is a model capable of aggregating talent, leveraging ingenuity while reducing the costs and time formerly needed to solve problems". One of the most recent works is by Pedersen et al. (2013, p. 585) who defined crowdsourcing as: "A collaboration model enabled by people-centric web technologies to solve individual, organisational, and societal problems using a dynamically formed crowd of interested people who respond to an open call for participation."

| Reference | Scope of review | Result |
|---|---|---|
| (Tripathi et al., 2014) | Crowdsourcing papers in the top 11 IS journals. | Typology of the types of crowdsourcing practiced and researched, and the types of potential problems crowdsourced by organizations. |
| (Ranard et al., 2014) | Peer reviewed literature that used crowdsourcing for health research. | Four distinct types of crowdsourcing tasks. |
| (Hetmank, 2013) | Papers about crowdsourcing system in peer-reviewed conference proceedings and journal papers since 2006. | 17 definitions of crowdsourcing systems were found and categorized into four perspectives |
| (Tarrell et al., 2013) | Crowdsourcing papers in the top 11 IS journals. | Analysis of keywords |
| (Pedersen et al., 2013) | Crowdsourcing papers in the top 11 IS journals. | A conceptual model of crowdsourcing |

*Table 1 Previous review papers on the crowdsourcing literature*

The research on different aspects of crowdsourcing has been started before development of the term. For example Brändle (2005) and Lin (2004) studied the effect of increase in the number of contributors on the quality of Wikipedia articles. Bryant et al. (2005) also used "activity theory" to describe a new paradigm for collaborative systems in which many people collaborate with each other to produce the final product.

However since 2006, the attention to this area has noticeably increased. The study of elite publications by Tarrell et al. (2013) indicates that number of publications in the area in 2012 has been 5 times more than this number in 2007. Pedersen et al. (2013) performed another crowdsourcing research classification work which categorized the crowdsourcing research into 6 groups which are: problem, process, technology, governance, people, and outcome.

In spite of the extensive literature on the crowdsourcing model in different areas, Pedersen et al. (2013) confirmed that the reviewed literature contains no comprehensive model in the process category. For this reason they extended the literature to other areas such as collaboration patterns (de Vreede and Briggs, 2005; Farooq et al., 2009) and called for comprehensive reviews and future research in this area. Previous case studies and research on the implementation of the crowdsourcing approach, however, have mentioned various activities as part of crowdsourcing projects. This study aims at synthesising these activities in the previous studies and developing a generic framework based on them. Table 1 lists some other review papers on the crowdsourcing research.

# 3 Research Method

The systematic literature review is a methodical way to identify, evaluate, and interpret the available empirical studies conducted on a topic, research question, or a phenomenon of interest (Kitchenham, 2004). For the purpose of the current study, we used guidelines provided by Kitchenham and Charters (2007) which involve five steps: (1) identify resources; (2) study selection; (3) data extraction; (4) data synthesis; and (5) write-up study as a report. The detailed process of selecting and reviewing the papers is depicted in Figure 1.

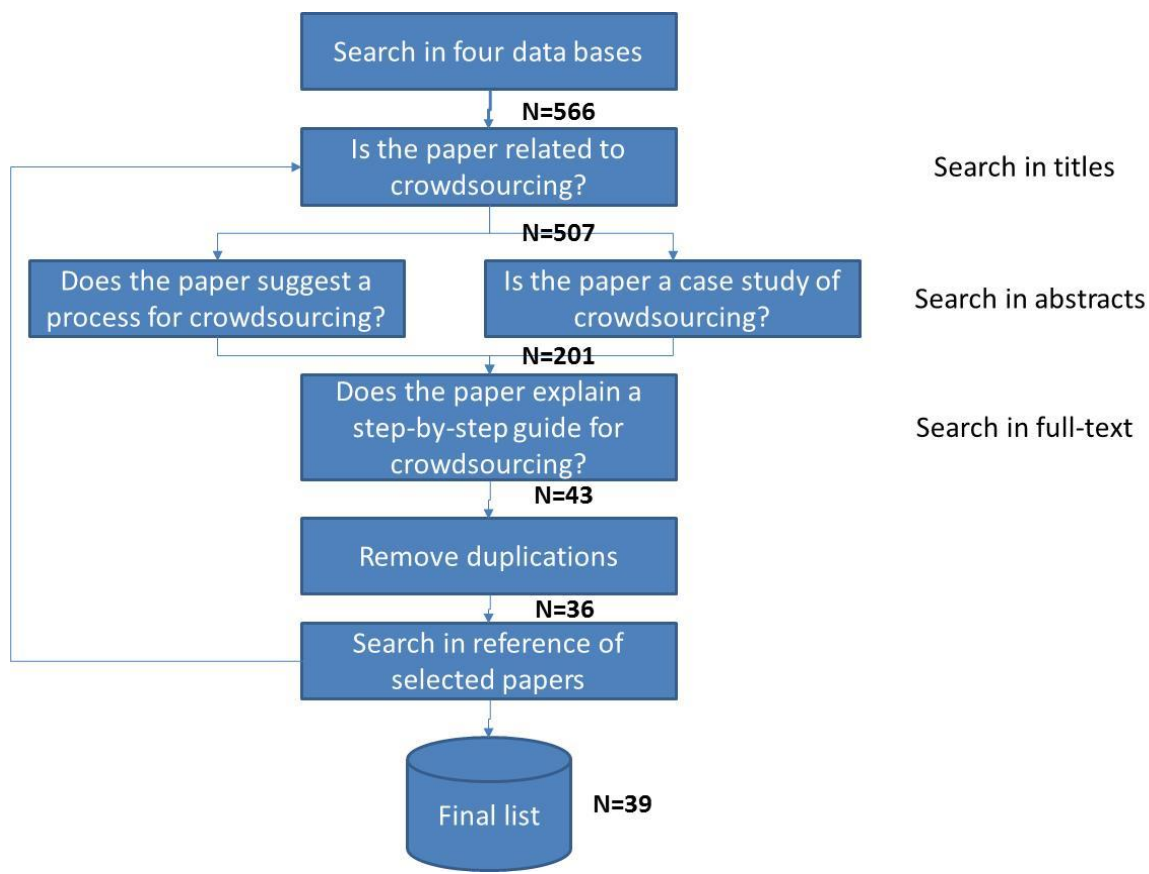

*Figure 1. Stages of Research Methodology*

To follow these steps the review started with searching the keywords in four scientific databases. The initial search resulted in 566 research papers. Irrelevant papers were then excluded during reviewing titles, abstracts and full-text papers. After in-depth study of the papers we arrived to final list of 39 papers and performed our analysis and classification based on those papers.

We tried to search the most well-known and comprehensive scientific databases in the field of IS for peer-reviewed publications. First of all we searched Scopus database which is recommended as a comprehensive source of scientific publications (Falagas et al., 2008; Meho and Yang, 2007) and indexes papers of many publishers like Elsevier, Emerald, and IEEE. We also searched three other famous databases in social science, management and IS areas which are Business Source Premier, ProQuest, and Association for IS electronic library. Finally in order to avoid overlooking any relevant paper, we checked the reference section of the selected papers in the final pool of research and searched the relevant papers in other databases. Search for peer-reviewed references in the well-known journals and conferences was the approach used by the current study to avoid researcher bias in the review process.

We searched for the following keywords on title, keywords and abstract of papers: (crowdsourcing OR "crowd sourcing" OR crowdsource OR crowd-based OR "Collective Intelligence") AND (Mechanism OR Process OR Procedure OR "case study" OR method OR step OR design OR framework OR plan OR phase). Table 2 indicates the number of retrieved research papers form each database in the final research pool.

| Database | Initial search | Final pool |
|---|---|---|
| Scopus | 363 | 24 |
| Business Source Premier | 69 | 2 |
| ProQuest | 87 | 5 |
| AISEL | 47 | 4 |
| Others | - | 4 |
| Total | 566 | 39 |

*Table 2. Number of papers from each database*

The initial search for the above phrases resulted in 566 papers. We then read the titles and abstracts and excluded irrelevant papers. After these rounds, the research pool decreased to 507 papers and then in another round we referred to the full texts to formulate a first list of 43 papers. While the selected four databases have some overlaps, we checked for duplicated papers and removed 8 papers. To make a second list, we verified the relevance of the sources used in those papers in order to find related studies. We found 38 studies referred to in those 35 papers. Then followed the above mentioned steps for these new 35 papers and found 4 new papers to our final list. We conducted the analysis based on a final list of 39 papers comprising the 35 papers of the first list and the 4 papers of the second.

In both shortlisting methods, we first excluded those papers that were not related to the topic of our research (crowdsourcing). In the next step, while reviewing abstracts, we excluded papers which were not related to crowdsourcing process or a case study on crowdsourcing implementation. Finally we read the remaining papers in full text to form the final set of papers which are the subject of our analysis. The main criterion for selecting relevant papers was their focus on processes and step by step approaches for crowdsourcing.

After compiling the final list of research papers, we started the analysis phase. We first paid attention to the section in which the process for crowdsourcing was provided. We then performed an in-depth analysis on the content of those papers and developed a general framework which covered most of the implementation activities in the processes. Finally synthesis of the collected data helped the research to develop a general process for crowdsourcing. Figure 1 illustrates the process of inclusion / exclusion.

## 4 Results

The final set of 39 papers formed the basis of the results described in upcoming sections. Figure 2 demonstrates number of papers in each year. As illustrated in this figure, crowdsourcing process has entered the literature in 2009, and except 2014 in which the current study is conducted, annual number of papers have remarkably increased between 2009 and 2013.

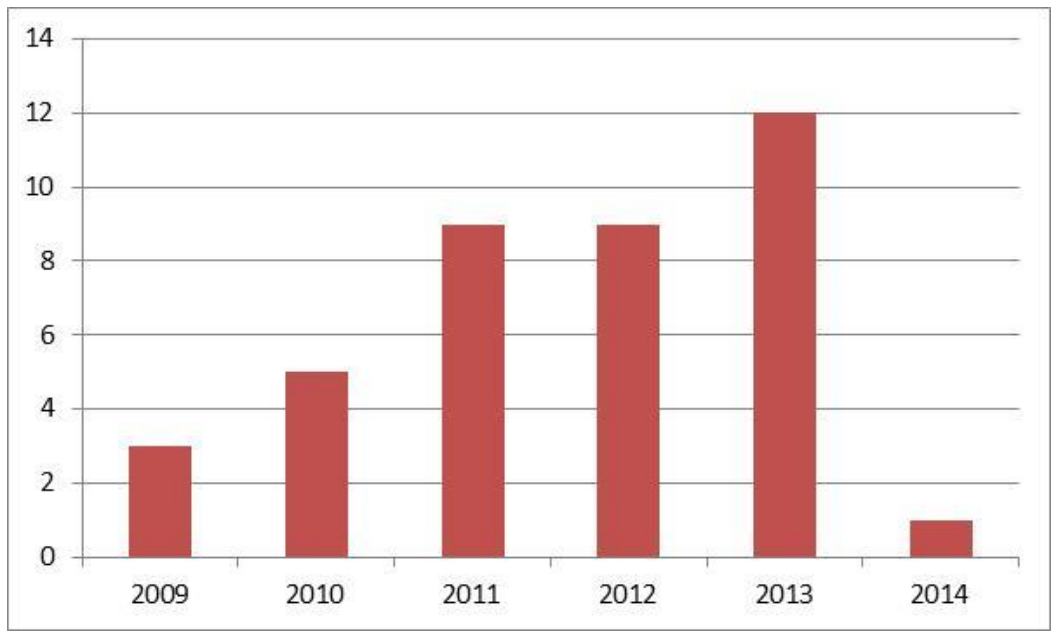

*Figure 2. Distribution of the studies*

During the review of different phases in the literature, we identified several sets of phases, and then, by adopting them with activities in different papers, we merged some phases, divided others into different phases, or renamed them. After several iterations, we finally arrived at our final set of phases which covered activities in all of the reviewed papers. Figure 3 illustrates the process for data analysis. Table 3 shows the titles we devised during several iterations for crowdsourcing.

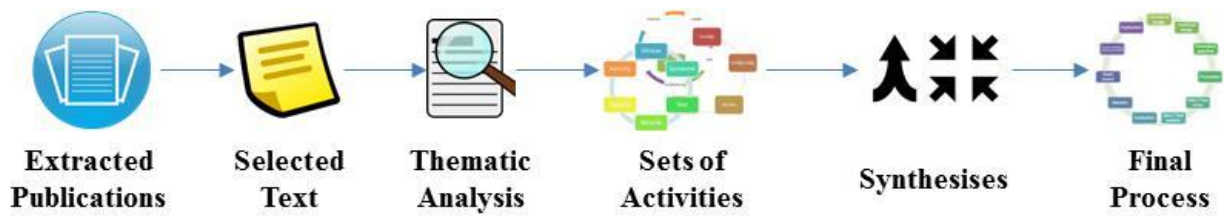

*Figure 3. Data Analysis Process*

| Sets of terms | Proposed Phases |
| --- | --- |
| 1st set | Pre-implementation →Implementation →Post-Implementation |
| 2nd set | Conceptual design → Technical design → Participant selection → Promotion → Technical development → Idea / Task Entry → Idea / Task revision → Evaluation → Grant award → Implement |
| 3rd set (final set) | Conceptual design → Technical design → Participant selection → Communication → Idea / Task Entry → Idea / Task revision → Evaluation → Monitor → Grant award → Process evaluation and documentation → Implement |

*Table 3. Different Iterations of Thematic Analysis*

Comparing our framework with the processes which we found in the literature review, some of them do not provide any suggestions for activities which can imply to one phase of suggested process model or mention activities which are (compared to the proposed process model) related to different phases. These mappings are performed in varying ways: sometimes one activity is assigned to one phase only, or alternatively, a number of different activities are assigned to one phase. Figure 4 illustrates the results of the synthesis of findings as a process model.

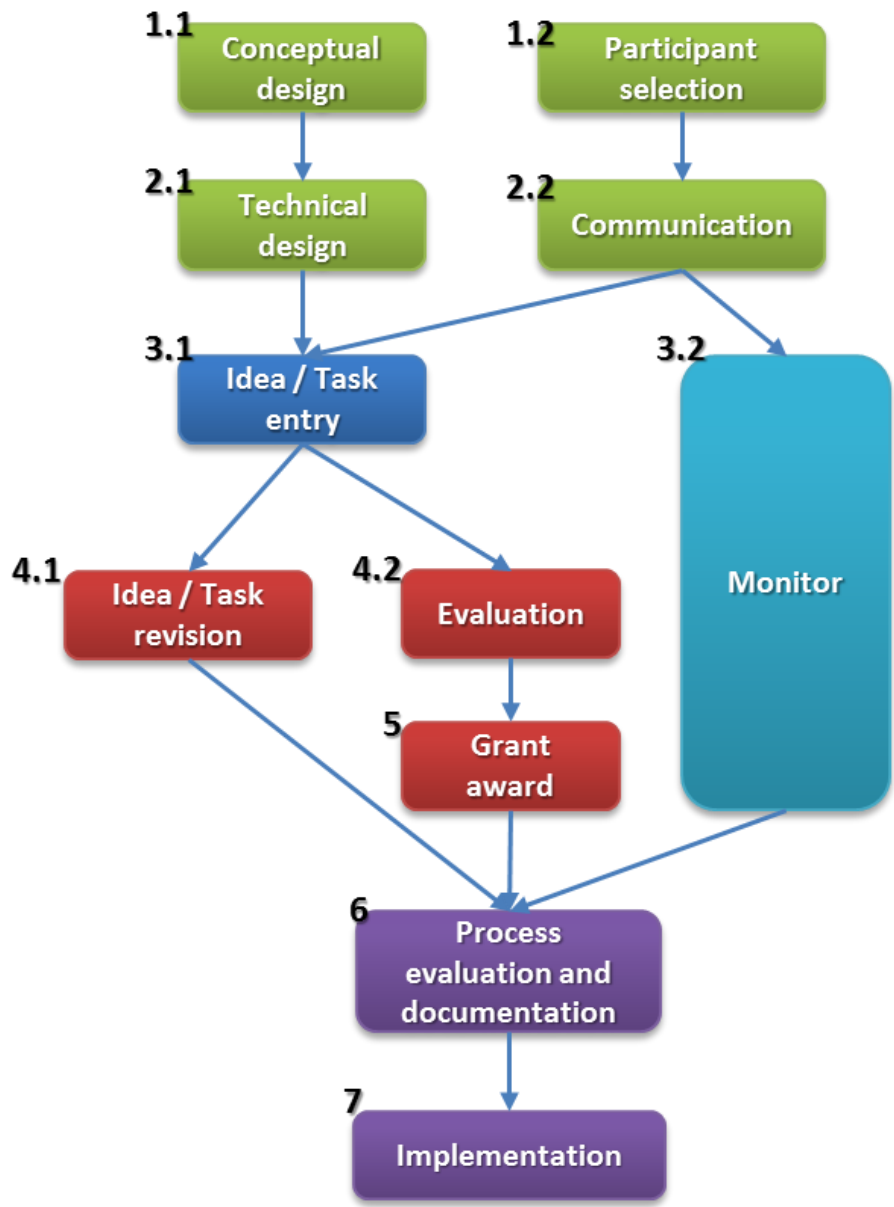

*Figure 4. Crowdsourcing Process Model*

*1.1 Conceptual design* phase covers activities which should be designed before the actual start of the project. This may include: definition of tasks which should be performed through crowdsourcing model (Anderson, 2011; Nguyen et al., 2013; Schulze et al., 2012), taking strategic decisions (Sutherlin, 2013), define goals (Lykourentzou et al., 2013), and define authorities and collaboration patterns (Bücheler and Sieg, 2011; Kuehn et al., 2011). The output of this phase could be a detailed plan for the activities which are going to be performed through crowdsourcing, a detail plan on time and people who should participate, and also manager's endorsement of the crowdsourcing project.

*1.2. Participant selection* phase deals with identification of crowd or "individuals who participate in the crowdsourcing problem (Pedersen et al., 2013)". This could be through selection of people inside and outside of the organisation (Bücheler and Sieg, 2011; Geiger et al., 2011; Park et al., 2013) contacting them (Chen and Liu, 2012; Hildebrand et al., 2013; Lorenzi et al., 2013), and in some cases performing some tests to select the crowd (Rossen and Lok, 2012; Stolee and Elbaum, 2010).

2.1. Activities in *technical design* phase answer the question: "how crowdsourcing should be performed?" and the development team decide to whether using available platforms (Costa-jussà et al., 2014; Schulze et al., 2012; Stolee and Elbaum, 2010) or develop a new platform for crowdsourcing (Liu et al., 2012; Park et al., 2013) and how to do this. Details on how platforms should work are determined in this phase as well and at the end of this phase, the crowdsourcing platform would be ready to work as an output of the performed activities.

2.2. In the *communication* stage, the organiser team invites the crowd to participate in the crowdsourcing project. While the number of submissions is recognised as an important factor which can affect the success of the crowdsourcing project (Walter and Back, 2011), this should be considered as an important phase in the project. A number of tools have been suggested in the literature for promoting crowdsourcing projects. Some of them are: open calls and advertisement (Potter et al., 2010; Wexler, 2011), direct correspondence with the selected crowd (Naparat and Finnegan, 2013; Ren, 2011), and providing training and workshops (Ebner et al., 2009).

*3.1 Idea / task entry* phase starts when the crowd start to interact with the crowdsourcing system and do their jobs in form of entering ideas (Degen, 2009; von Briel and Schneider, 2012; Vuori, 2012) or performing a task (Liu et al., 2012; Nguyen et al., 2013; Schulze et al., 2012).

3.2 There are a number of tasks which are required to be performed by the organising team during the implementation of the project to make sure the process is going on in the desired way. In this study these tasks have been categorised as *monitor* phase. Some of these activities are: coordination of the crowd (Alam and Campbell, 2013; Naparat and Finnegan, 2013), manage concurrency, manage time (Müller et al., 2010), and sanction of inappropriate entries (Alam and Campbell, 2013).

4.1 In most of the crowdsourcing projects the responsibility of the selected crowd is not only to enter their inputs to the system, but also rank, filter, revise, and comment on others inputs which can be categorised as *idea / task revision* phase. The main purpose of this phase is to remove any possible error from inputs which are already submitted (Rossen and Lok, 2012) and synthesis or integrate them (Naparat and Finnegan, 2013). More than the available crowd, some crowdsourcing systems have authorised experts in the organisation for correction, filtering, or revision of the entries (Wexler, 2011).

4.2 While most crowdsourcing systems use different types of incentives to motivate the crowd and also in order to select the most useful inputs in the system, *evaluation* of entries is needed as part of the process. This evaluation may take place simultaneously with the previous phase or independent of other phases. It can also be performed by selected experts in the field (Chen and Liu, 2012; Ebner et al., 2009).

*5. Granting the award* is the next phase in the process. Crowdsourcing problems are generally categorised in two groups of competition and collaboration or outcome-based and contribution-based (Markus et al., 2002) and this phase is more applicable to the first category. Kaufmann et al. (2011) also mentioned two categories of "extrinsic motivation" (immediate payoff, delayed payoff, and social payoff) and "intrinsic motivation" (fun, enjoyment, and social interaction) for participation in crowdsourcing. In order to increase motivation for future practices of crowdsourcing, granting the rewards in all types has been mentioned as part of the crowdsourcing process in the literature.

*6. Process evaluation and documentation* of the lesson learned from the crowdsourcing project has been suggested as an important part of the crowdsourcing process. According to the literature, activities in this phase may include: knowledge documentation and management (Anderson, 2011; Vuori, 2012), arrange future practices (Wexler, 2011), and plan for retaining the crowd (Ren, 2011).

7. Future actions to *implement* the results of the crowdsourcing project have also been cited as part of the process in some references. It contains diffusion and presentation of the ideas (Rossen and Lok, 2012; von Briel and Schneider, 2012) and implementation or commercialization of the result (Degen, 2009; Potter et al., 2010). Table 4 provides a brief introduction to each phase in the developed process model.

| Crowdsourcing phase | Definition | Possible activities in the literature |
|---|---|---|
| *Conceptual design* | Activities which should be performed before start of the technical development of the project | Describe the task, Design motivation system, Plan the job |
| *Technical* | Design and development of the | Platform selection, Platform design |

| Crowdsourcing phase | Definition | Possible activities in the literature |
|---|---|---|
| *design* | crowdsourcing platform | and development |
| *Participant selection* | Selection of the crowd who will participate in the crowdsourcing task | User selection, Team Formation |
| Communication | Communication with the selected crowd | Contact the crowd, Crowdsourcer broadcasting, Communication |
| Idea / Task entry | The crowd start their interaction with the system | Task choice, Idea generation, Collect inputs from crowd, Execute job, Find solution |
| Idea / Task revision | Rank, filter, revise, and comment on the crowd's inputs | Conversion, Clarification, Aggregation of contributions, Refine task, Collaborative, Solution synthesis |
| Evaluation | Check the appropriateness of the inputs from the crowd | Result / input evaluation, Analyse the result, Competitive and judging process |
| Monitor | Organization of the team during the implementation | Coordination, Manage concurrency / input and output / time, Workflow Management |
| Grant award | Identification of the best entry and awarding related incentive | Prize for winning entries, Idea awards ceremony, Reward |
| Process evaluation and documentation | Review the project and document the lesson learned for future improvements | Knowledge retrieval, Mentoring session, Knowledge capture, Knowledge evaluation, Decide on future crowdsourcing arrangements, Evaluation of the project, Post-competition, Train models |
| Implementation | Implement the results of the crowdsourcing | Collaborative discussion, Presentation of result, Collective action, Results and Analysis, Post-competition, Implementation Actions |

*Table 4. Phases in the Final Framework*

## 5  Discussion

This paper reviewed the available methods and case studies of crowdsourcing and base on an in-depth analysis of their recommendations, developed a comprehensive process model that covers most of the available result. Figure 5 illustrates the frequency of citation for each phase in the literature. As can be seen in the graph after the idea / task entry, conceptual design and evaluation phases have been subject of attention in most of the papers. Also monitoring, process evaluation, and implementation phases have been less studied as part of the crowdsourcing process in the related literature. Current study calls for more attention to these phases in future research and case studies.

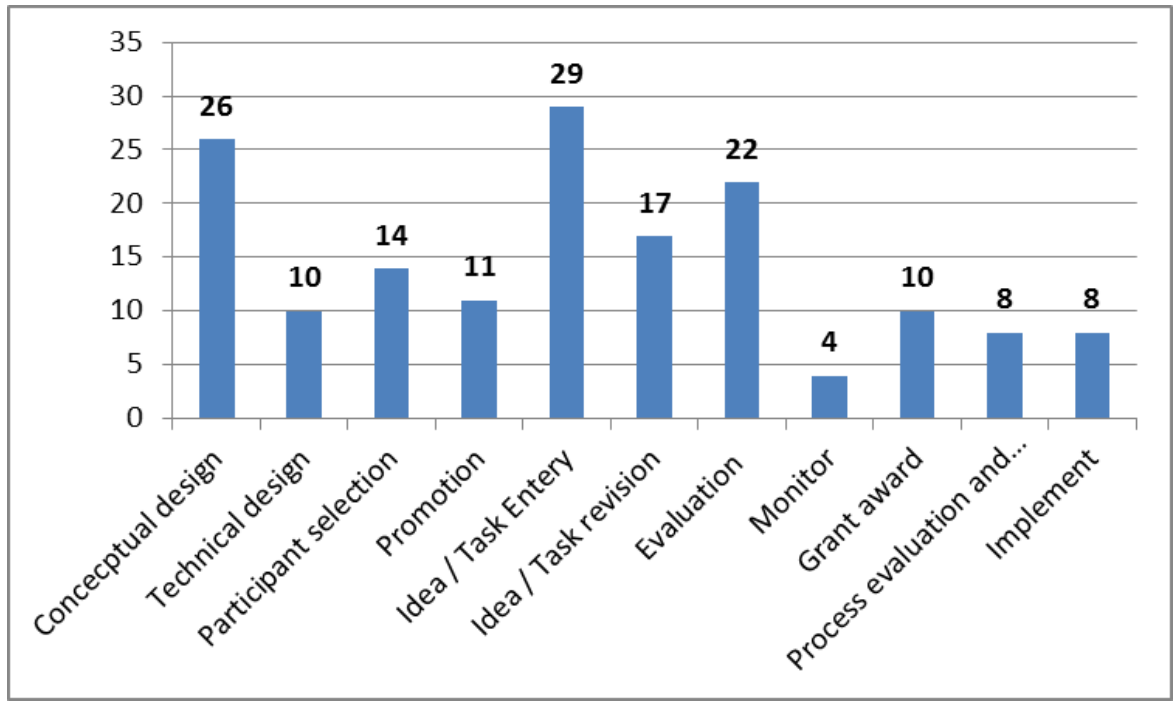

*Figure 5. Number of Citation in the literature for each Phase*

The current study provides practitioners with guidelines for comparing and selecting an appropriate process of crowdsourcing. Different organisations, depending on the context of their business and their time and budget limitations, can select a process which best fits their goals. Those who have already selected a process can also better understand the shortcomings of their current method and could perform alternative activities to ameliorate those shortcomings. The framework may also help consultant companies to select, develop or modify the crowdsourcing process they suggest for their customers.

The processes introduced in this paper are diverse in details (from two to six activities) and context. For this reason a variety of practitioners (including CEOs, CIOs, consultants, and IS/IT personnel) in divers organisations may benefit from the current review. The provided framework itself can be used as a process model for future practices of crowdsourcing.

## 6   Limitation

This paper is a first step toward developing a process model which will be developed through design science research method and for this reason it lacks empirical evaluation. Moreover, the used systematic literature review method obviously entails limitations and restrictions such as: generality of work and lack of empirical studies on the suitability of the general framework for different contexts. Future studies will refine the framework and adopt it within various contexts.

Companies of very diverse scales from global companies such as IBM (Stewart et al., 2009) to small start-up businesses are currently utilizing crowdsourcing. For this reason the proposed process model should be tailored in most cases considering the contextual elements of each business including size, business type, strategic directions, crowd workers, and used platform. Especially start-ups which plan to use pre-developed platforms (such as Mechanical Turk) are most limited in adopting some activities in the framework. However this process model can help them in selecting the best available activities for their project.

## 7   References

Alam, S. L., and Campbell, J. (2013). "Role of Relational Mechanisms in Crowdsourcing Governance: An Interpretive Analysis.").